%% file: paper.tex
\documentclass[openany,pdftex,twocolumn,epjc3]{svjour3}
\usepackage{./sty/ply}
\usepackage[normalem]{ulem} 

\journalname{Eur. Phys. J. C}
\begin{document}

\title{Measurement of proton light yield of water-based liquid scintillator}
\input{inc/authors}
\date{Date: \today}
\maketitle

\input{inc/abstract}

\input{inc/introduction}

\input{inc/setup}

\input{inc/analysis}

\input{inc/results}

\input{inc/conclusion}

\input{inc/acknowledgements}

\bibliographystyle{spphys}
\bibliography{wbls-ply.bib}

\end{document}

%% file: inc/authors.tex

\newcommand{\ucbphy}{Physics Department, University of California at Berkeley, Berkeley, CA 94720, USA}
\newcommand{\lbnlnsd}{Nuclear Science Division, Lawrence Berkeley National Laboratory, Berkeley, CA 94720, USA}
\newcommand{\ucbeng}{Department of Nuclear Engineering, University of California at Berkeley, Berkeley, CA 94720, USA}
\newcommand{\bnlchm}{Chemistry Division, Brookhaven National Laboratory, Upton, NY 11973, USA}
\newcommand{\ejc}{ejc3@berkeley.edu}
\newcommand{\contactline}{mailto:\ejc?subject=EPJC\%20WbLS\%20PLY}

\author{{E. J. Callaghan}   \thanksref{ucbphy,lbnl,contact}
    \and {B. L. Goldblum}   \thanksref{lbnl,ucbeng}
    \and {J. A. Brown}      \thanksref{ucbeng}
    \and {T. A. Laplace}    \thanksref{ucbeng}
    \and \\ 
         {J. J. Manfredi}   \thanksref{ucbeng,afit}
    \and {M. Yeh} \thanksref{bnlchm}
    \and {G. D. Orebi Gann} \thanksref{ucbphy,lbnl}
}
\institute{\ucbphy\label{ucbphy}
    \and \lbnlnsd\label{lbnl}
    \and \ucbeng\label{ucbeng}
    \and \bnlchm\label{bnlchm}
}
\thankstext{contact}{email: \href{\contactline}{\texttt{\ejc}}}
\thankstext{afit}{Present address: Department of Engineering Physics, Air Force Institute of Technology, Wright-Patterson Air Force Base, OH 45433 USA.}

%% file: inc/abstract.tex

\begin{abstract}

    The proton light yield of liquid scintillators is an important property in the context of their use in large-scale neutrino experiments, with direct implications for neutrino-proton scattering measurements and the discrimination of fast neutrons from inverse $\beta$-decay coincidence signals. This work presents the first measurement of the proton light yield of a water-based liquid scintillator (WbLS) formulated from 5\% linear alkyl benzene (LAB), at energies below 20\;\MeV{}, as well as a measurement of the proton light yield of a pure LAB + 2\;\gram\per\liter{} 2,5-diphenyloxazole (PPO) mixture (LABPPO). The measurements were performed using a double time-of-flight method and a pulsed neutron beam from the 88-Inch Cyclotron at Lawrence Berkeley National Laboratory. The proton light yields were measured relative to that of a 477\;\keV{} electron. The relative proton light yield of WbLS was approximately 3.8\% lower than that of LABPPO, itself exhibiting a relative proton light yield $15-20\%$ higher than previous measurements of an analogous anoxic sample. The observed quenching is not compatible with the Birks model for either material, but is well described with the addition of Chou's bimolecular quenching term.

\end{abstract}

%% file: inc/introduction.tex

\section{Introduction}
\label{sec:intro}

    Neutrinos provide a gateway to improved understanding of basic physics, though their fundamental nature remains unknown. Liquid scintillators have been a mainstay for experimental neutrino physics, from the Cd-loaded toluene medium employed by Reines and Cowen \cite{Reines1956} to the linear alkylbenzene (LAB) based systems of today \cite{An2012,Andringa2016}. While liquid scintillators demonstrate high efficiency for the conversion of particle kinetic energy into detectable light, the isotropic emission of scintillation photons makes generic reconstruction of neutrino directionality notoriously difficult.

    Water-based liquid scintillator (WbLS) \cite{Yeh2011} has emerged as a versatile detection medium for large-volume neutrino detectors, capable of leveraging both the Cherenkov and scintillation light, and is a candidate material to be deployed in upcoming liquid-phase detectors including ANNIE \cite{ANNIE}, AIT-NEO \cite{AITNEO}, and {\sc Theia} \cite{Askins2020}. It is composed of organic liquid scintillator encapsulated in micelles and dispersed in a water solvent, which has the advantage of providing increased light yield relative to traditional water-based detectors with only nominal increase in cost, while retaining a relatively clear Cherenkov signal. Prior measurements of the scintillation emission spectrum, light yield, and temporal response, as well as demonstrated separation between Cherenkov and scintillation photon populations \cite{Bignell2015,Onken2020,Caravaca2020} offer the possibility of vertex reconstruction comparable to that achieved with pure liquid scintillator but with improved directional sensitivity \cite{Land2021}. Given this, WbLS holds promise for enabling new hybrid neutrino detector design concepts, which admit robust directional reconstruction with lower detection thresholds.

    The advantages offered by WbLS extend the reach of neutrino detectors to several fundamental science goals \cite{Askins2020}. The relatively low cost allows for the construction of larger detectors, with the low energy threshold enabling large-scale searches for neutrinos from the Diffuse Supernova Neutrino Background (DSNB) and proton decay, for example, and directional reconstruction capabilities offering enhanced measurements of low energy solar neutrinos. The relative abundance of Cherenkov light allows for high-precision ring imaging, which improves particle-identification capability, improving sensitivity in e.g., long-baseline oscillation measurements.
  
    In addition to the basic science applications, recent advances in antineutrino physics technologies have motivated considerable interest in WbLS for neutrino-based reactor monitoring. Through measurement of the fission neutrino signal from a nuclear reactor, it is theoretically possible to discern the reactor power level and isotopic composition of the fuel, important proliferation indicators for nuclear security applications \cite{Bernstein2020}. Given the low neutrino interaction cross section, the accurate prediction of background signals arising from ambient radioactivity is critical. For example, fast neutrons from cosmogenic muon interactions represent an important source of background for inverse $\beta$-decay (IBD) measurements. Internal radioactive contaminants, e.g., neutrons produced via the $^{13}$C($\alpha,n$)$^{16}$O reaction, may represent additional background contributors \cite{Yoshida2010}. As fast neutrons primarily generate light in scintillating media via $np$ elastic scattering before capturing, measurement of the proton light yield of the WbLS is essential in distinguishing neutron interactions from true IBD events.

    This work presents the first measurement of the proton light yield of WbLS, loaded at the level of 5\% scintillator concentration. A measurement of LAB with 2 g/L 2,5-diphenyloxazole (PPO), henceforce denoted LABPPO, was also conducted to serve as a fiducial reference. \refsec{setup} provides a description of the experimental setup and associated electronics configuration. In \refsec{analysis}, the analytic methods are described, including the calibrations of the electronics and energy reconstruction, the extraction of the proton light yield, and tests of quenching model compatibility. \refsec{results} presents the measured proton light yield (PLY) relations of WbLS and LABPPO in the energy range of 2 to 20\;\MeV{} along with ionization quenching model fits. Concluding remarks are given in \refsec{conclusion}.

%% file: inc/setup.tex

\section{Experimental setup}
\label{sec:setup}

\begin{figure}
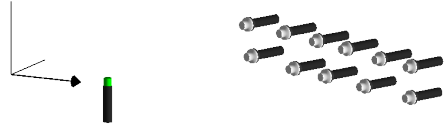

	\centering
  	\img{0.45\tw}{setup/geant4-schematic-transparent.png}
	\caption{Experimental setup for proton light yield measurements. The neutron beam travels along the axis designated with an arrow to the target scintillator cell, shown in green. Eleven auxiliary detectors are positioned at forward scattering angles with respect to the incoming neutron beam.}
	\label{fig:expSetupFig}
\end{figure}

    A broad spectrum neutron beam was produced by impinging a 33\;\MeV{} $^{2}$H$^{+}$ beam onto a 3-mm-thick Be target at the 88-Inch Cyclotron at Lawrence Berkeley National Laboratory \cite{Harrig2018}. The LABPPO and WbLS samples to be characterized were independently placed in beam, about 7\;\meter{} downstream of the breakup target. Eleven auxiliary detectors, filled with EJ-309 \cite{EJ309}, an organic liquid scintillator with pulse-shape-discrimination (PSD) capabilities, were positioned out of beam to detect forward-scattered neutrons from the target scintillator. A schematic diagram of the experimental setup is shown in \reffig{expSetupFig}. The detector geometries employed for the two measurements are provided in \reftbl{geometry}. The geometry was established using laser-based coordinate measurements, assigning a 1\;\centi\meter{} uncertainty to each measurement except the $z$-position of the breakup target, which is known to 5\;\milli\meter{}.
    
\begin{table*}
    \centering
    \small
    \begin{tabular}{c c | c c | c c}
        \- &
            \- &
            \multicolumn{2}{c |}{
                LABPPO
            } &
            \multicolumn{2}{c}{
                WbLS
            } \\
        \- &
            Channel &
            Distance [\centi\meter{}] &
            Scattering angle [\degree] &
            Distance [\centi\meter{}] &
            Scattering angle [\degree] \\
        \hline
        Breakup to target &
            \- &
            $721.3 \pm{1.4}$ &
            -- &
            $716.6 \pm{1.4}$ &
            -- \\
        Target to &
            2 &
            $133.8 \pm{1.8}$ &
            $80.0 \pm{1.9}$ &
            $134.2 \pm{1.8}$ &
            $78.0 \pm{1.9}$ \\
        \- &
            3 &
            $131.7 \pm{2.1}$ &
            $65.0 \pm{2.0}$ &
            $133.2 \pm{2.1}$ &
            $63.1 \pm{2.0}$ \\
        \- &
            4 &
            $137.6 \pm{2.2}$ &
            $52.2 \pm{2.0}$ &
            $140.0 \pm{2.1}$ &
            $50.5 \pm{1.9}$ \\
        \- &
            5 &
            $148.1 \pm{2.2}$ &
            $41.9 \pm{1.9}$ &
            $151.1 \pm{2.1}$ &
            $40.4 \pm{1.8}$ \\
        \- &
            6 &
            $165.4 \pm{2.1}$ &
            $32.3 \pm{1.7}$ &
            $168.9 \pm{2.0}$ &
            $31.2 \pm{1.7}$ \\
        \- &
            7 &
            $184.9 \pm{2.0}$ &
            $25.1 \pm{1.6}$ &
            $188.7 \pm{1.9}$ &
            $24.2 \pm{1.5}$ \\
        \- &
            9 &
            $133.0 \pm{1.7}$ &
            $78.1 \pm{1.9}$ &
            $134.1 \pm{1.6}$ &
            $76.2 \pm{2.0}$ \\
        \- &
            12 &
            $132.7 \pm{1.9}$ &
            $61.4 \pm{2.0}$ &
            $135.1 \pm{1.9}$ &
            $59.7 \pm{2.0}$ \\
        \- &
            13 &
            $139.6 \pm{2.0}$ &
            $48.7 \pm{2.0}$ &
            $142.7 \pm{1.9}$ &
            $47.3 \pm{1.9}$ \\
        \- &
            14 &
            $156.2 \pm{2.0}$ &
            $35.9 \pm{1.8}$ &
            $160.0 \pm{1.9}$ &
            $34.9 \pm{1.8}$ \\
        \- &
            15 &
            $183.7 \pm{1.8}$ &
            $24.4 \pm{1.6}$ &
            $187.9 \pm{1.8}$ &
            $23.8 \pm{1.5}$ \\
            \hline
    \end{tabular}
    \caption{Distances between various experimental apparatus, and nominal scattering angles associated with each auxiliary detector.}
    \label{tbl:geometry}
\end{table*}

\begin{figure*}
    \centering
    \includegraphics[width=0.35\tw,angle=0]{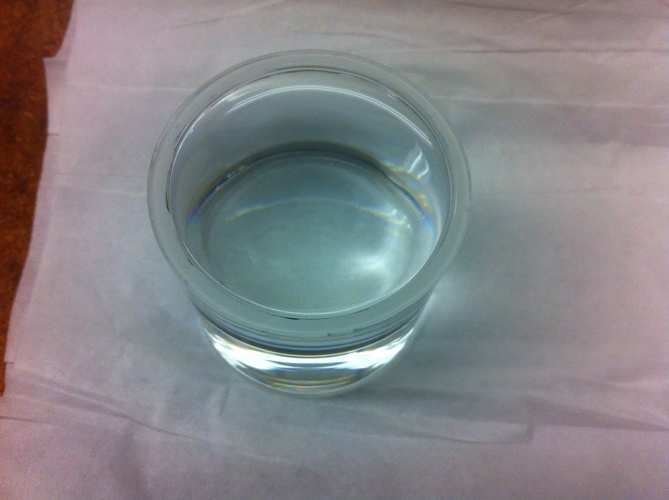}
    \hspace{0.1\tw}
    \includegraphics[width=0.35\tw,angle=0]{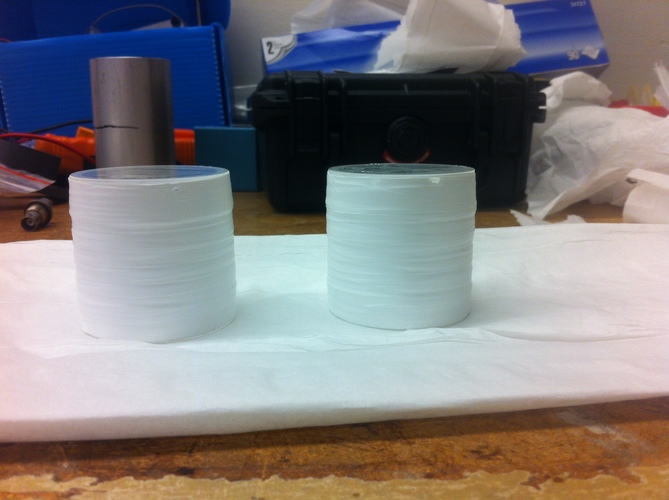}
    \caption{(Left) Sealed target cell containing LABPPO before wrapping with PTFE tape. (Right) Both target cells after wrapping in PTFE.}
    \label{fig:cells}
\end{figure*}

    The LABPPO and WbLS target scintillators were contained in cylindrical quartz crucibles, of dimensions 50\;\milli\meter{} diameter by 50\;\milli\meter{} tall and 1\;\milli\meter{} in wall thickness. A quartz disk of the same thickness was used to seal the open face using a two-part epoxy. The side wall and sealed top of the cells were wrapped in no less than 10 layers of polytetrafluoroethylene (PTFE) tape to improve internal reflectivity, and thus light collection. The remaining transparent face was optically coupled to a Hamamatsu H1949-51 photomultiplier tube (PMT) using EJ-550 silicone grease. The sealed cells, both before and after wrapping with PTFE, are shown in \reffig{cells}.

    The scintillator cells of the auxiliary detectors were right cylinders of diameter and height 50.8\;\milli\meter{}, constructed of a thin aluminum housing and filled with EJ-309 \cite{EJ309}, and were each coupled to a PMT via a borosilicate glass window and EJ-550 silicone grease. All PMTs used in these measurements were obtained from Hamamatsu Photonics (either Type No. 1949-50 or 1949-51), and were negatively biased using either a CAEN R1470ET or CAEN NDT1470 power supply.

    For each sample, data were acquired over a period of approximately 11 hours, with a beam current of approximately 55\;\nano\ampere{}. The data acquisition system triggered on a coincidence between the target PMT and any of the auxiliary detectors within a 400\;\nano\second{} coincidence window. Upon triggering, digital waveforms of a total length of 800\;\nano\second{} from all channels, as well as a waveform digitizing a sinusoidal RF control signal provided by cyclotron operations, were recorded using a CAEN V1730 500 MS/s digitizer. The scintillator signal timing was determined using the CAEN digital constant fraction discrimination algorithm, with a 75\% fraction and a 4\;\nano\second{} delay. The timing pickoff for the cyclotron RF signal was determined using leading-edge discrimination.

%% file: inc/analysis.tex

\section{Analysis methods}
\label{sec:analysis}

Waveforms in the target detectors were integrated for 140\;\nano\second{} to ensure collection of $\geq95\%$ of the observed charge. For the auxiliary detectors, waveforms were integrated for 300\;\nano\second{} to provide an integrated charge, and a PSD-metric was obtained by calculating the ratio of the integrated charge of the prompt region corresponding to the first 30\;\nano\second{} of the waveform, to the delayed region between 30\;\nano\second{} and 260\;\nano\second{} from the start of the waveform, providing good separation between $\gamma$-ray and neutron signals for high charge producing events. For coincident events, the high-level observables are the integrated charge and timing for the target and auxiliary detectors, a PSD-metric for the auxiliary detector, and a timestamp corresponding to the cyclotron RF signal. To measure the proton light yield (PLY) as a function of energy, a conversion between charge and light must be established, $\gamma$ and neutron interactions distinguished, and the energy deposited by neutron interactions reconstructed from the available timing and geometric information. The methods employed herein were originally introduced in\;\cite{Brown2018,BrownThesis} and are further detailed below.
    
\subsection{PMT linearity correction}
\label{sec:pmt-linearity}

\begin{figure}
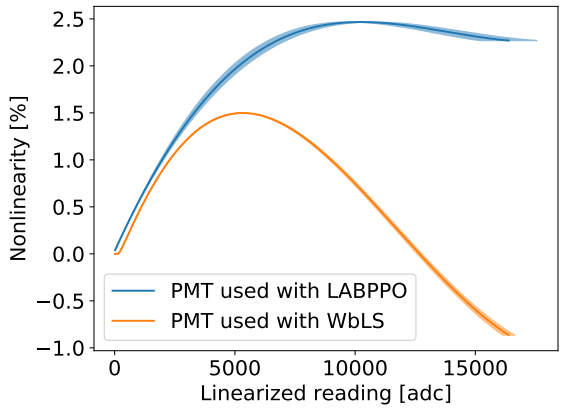

    \centering
    \img{0.45\tw}{analysis/nonlinearity-degree-4-overlay-20x-uncertainties.pdf}
    \caption{Deviation of output current from linear operation of the PMTs used in this measurement, as biased during runtime operations. The blue and orange curves correspond to the PMTs mounted to the LABPPO and WbLS samples, respectively. The abscissa spans the full scale range of the employed digitizer. The uncertainty bands are scaled by a factor of 20 for visualization.}
    \label{fig:pmt-nonlinearity}
\end{figure}

   A nonlinearity correction for the two PMTs coupled to the measurement samples was performed using the method of Friend et al.\ \cite{Friend2011}. In brief, each PMT was placed in the view of two LEDs with peak wavelength 405\;\nano\meter{} \cite{LEDspec}, which were flashed both independently and in coincidence, thus recording the PMT response to two independent fluxes, as well as the response to the summed flux. By repeating this procedure over a range of fluxes spanning the range of the digitizer used in this measurement, the deviation from linear operation was computed. The measured nonlinearities, interpreted as quartic polynomials, are shown in \reffig{pmt-nonlinearity}. The nonlinearity correction was applied on a sample-by-sample basis to waveforms collected both during reference charge calibration and beam running.

\subsection{Reference charge calibration}
\label{sec:charge-calibration}
    To establish a measurement unit proportional to the number of scintillation photons, a reference charge is defined and serves as a calibration. In this work, the reference charge is that associated with a 477\;\keV{} electron, evaluated using the Compton edge of the 662\;\keV{} $\gamma$ ray following \isotope{Cs}{137} decay. Calibration data were collected using \isotope{Cs}{137} (662\;\keV{}) and \isotope{Bi}{207} (1.770\;\MeV{}) sealed sources, as well a \isotope{Na}{24} (2.754\;\MeV{}) source created by beam-activation of a sample of natural aluminum, placed at distances $\geq 10\;\centi\meter$ from the center of each target scintillator cell. Because beam operation was required to produce the sample of \isotope{Na}{24}, this source was not available before irradiation of the LABPPO sample. For each other source used with the LABPPO scintillator, and for all sources used with the WbLS, calibration data were taken both before and after irradiation.

    The charge associated with the Compton edge, or the Compton charge, was determined by fitting a model to the measured calibration data. The model consists of an electron energy deposition spectrum following $\gamma$-ray interactions in the scintillator, generated using the \texttt{GEANT4} simulation toolkit \cite{GEANT4}, convolved with a three-parameter system resolution function \cite{Dietze1982} as well as a power-law background term \cite{Laplace2020-EJ20x}. A linear charge response was applied to the experimental data to convert the measured charge in analog-to-digital converter (ADC, or adc) channels, $Q$, to that associated with a given electron recoil energy, $E$. The energy-charge relation is $E = a Q + b$, which assumes that the electron light yield is approximately linear in the energy range of interest, with $b$ accounting for potential nonlinearity at lower energies. The minimization was performed using the SIMPLEX and MIGRAD algorithms from the ROOT Minuit2 package \cite{ROOT}.

    For each target scintillator, the measured calibration data before and after neutron irradiation were fit with the corresponding charge model independently, with the offset term, $b$, fixed to zero. The resulting Compton charges are reported in \reftbl{charge-calibration} for each $\gamma$-ray source, along with the statistical uncertainty, determined from the parameter uncertainty on $a$, and systematic uncertainty stemming from the uncertainties in the background shape and electron light linearity, described in detail below.

\begin{figure*}[t!]
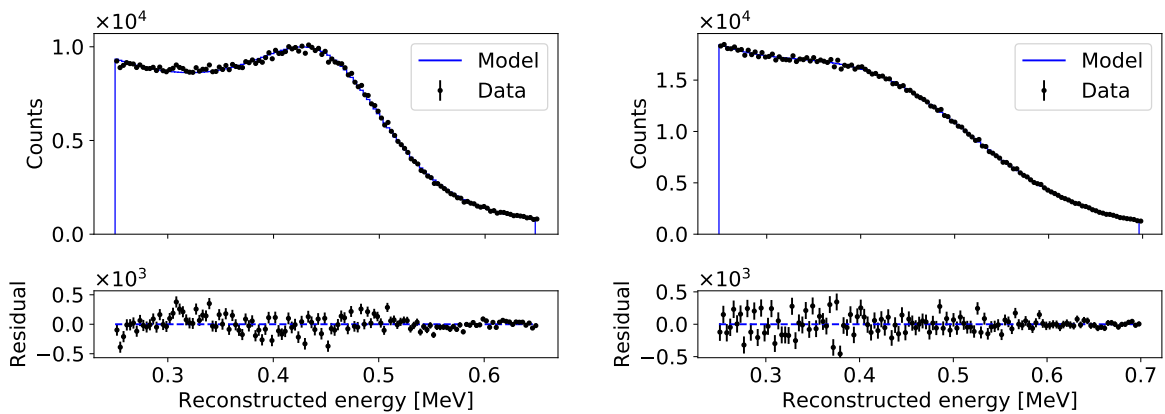

	\centering
	\img{0.45\tw}{analysis/labppo-137cs-fit-tlaplace.pdf}
	\img{0.45\tw}{analysis/wbls-137cs-fit-tlaplace.pdf}
	\caption{Best fit charge models compared to LABPPO (left) and WbLS (right) calibration data using a $^{137}$Cs source.}
	\label{fig:137cs-fits}
\end{figure*}

    The systematic uncertainty on the Compton charge is computed as the standard deviation of the Compton charge determined using all available combinations of pairs and triplets of calibration $\gamma$ rays. Simultaneous fits to multiple Compton edges were performed without any constraint on $b$, the value of which provides information about low-energy electron light nonlinearity. For LABPPO, $b = \pp{34.71 \pm 1.24}\;\keV$, and for the WbLS, $b = \pp{185.3 \pm 4.0}\;\keV$, where the uncertainty on $b$ is given by the standard deviation of the values obtained by fitting all available combinations of pairs and triplets. The difference in $b$ for LABPPO and the WbLS can be attributed, in part, to the larger relative contribution of Cherenkov-to-scintillation light in WbLS compared to LABPPO, and should be taken into account when comparing to quenching measurements obtained using different $\gamma$-ray sources for light calibration. The best-fit charge models are compared to the \isotope{Cs}{137} data in \reffig{137cs-fits}.

    The gain stability of the target PMTs was investigated by chronologically partitioning the full beam dataset for each scintillator into 10 distinct datasets and analyzing each separately. No systematic trends or significant fluctuations were observed in the PLY results. A strong ambient $\gamma$-ray background was present in the experimental hall, which was further exacerbated by irradiation of the hall and experimental apparatus during data collection, and can introduce bias in the determination of the Compton charge. A 1.8\% and 1.5\% difference in the estimated Compton charge extracted using the \isotope{Cs}{137} calibration data before and after in-beam irradiation for LABPPO and WbLS, respectively, is attributed to this variation in the background. Smaller variations were observed for the \isotope{Bi}{207} lines (0.1\% and -0.3\%) and the \isotope{Al}{24} data (0.6\%), which are in a higher energy region where the background contribution is less significant.

\begin{table*}
	\centering
	\begin{tabular}{c  c  c  c }
	& & \multicolumn{2}{c}{Compton charge [\adc{}]\ } \\
	Source & Compton edge energy [\keV{}] &  LABPPO & WbLS \\
	 \hline
	$^{137}$Cs & 477 & 2525.4 $\pm$ 1.3 $\pm$ 21.2 &
		2131.9 $\pm$ 2.6 $\pm$ 53.9 \\

	$^{207}$Bi & 1547 & 
	8617.0 $\pm$ 22.8 $\pm$ 13.7 & 
		9741.2 $\pm$ 82.8  $\pm$ 94.7\\
		
	$^{24}$Na & 2520 & 	
		14219.5 $\pm$ 9.3 $\pm$ 44.5 &
		16795.5 $\pm$ 15.7 $\pm$ 71.5\\
		\hline
	\end{tabular}
	\caption{Compton charges for both LABPPO and WbLS. The first uncertainty corresponds to the statistical uncertainty obtained from parameter fitting. The second uncertainty corresponds to the standard deviation of the Compton charge determined using simultaneous fits of multiple calibration spectra (i.e., all combinations of pairs and triplets).}
	\label{tbl:charge-calibration}
\end{table*}

\subsection{Auxiliary detector particle identification}
\label{sec:psd}

    The 11 auxiliary detectors located at forward scattering angles are filled with EJ-309 \cite{EJ309}, a commercial liquid scintillator with established particle-identification (PID) capabilities achieved via PSD, in this case exploiting that $\gamma$-ray pulses have a higher ratio of prompt to delayed light relative to neutron pulses. For each auxiliary detector, a constraint on the total charge collected is chosen to reject events in the low-charge region where the distributions of PSD values from pulses originating from neutron and $\gamma$-ray interactions overlap. These constraints are then imposed on beam data, after which the PSD metric, i.e., the ratio of delayed to prompt charge, is binned and fit with an empirical normal-plus-lognormal form, where the former term models the distribution of $\gamma$s and the latter neutrons. After performing the fit, an optimal PSD value for distinguishing between the two components is determined by minimizing the neutron contamination of $\gamma$ selection, with the resultant purity above 98\% for high-charge events.

\begin{figure*}[t!]
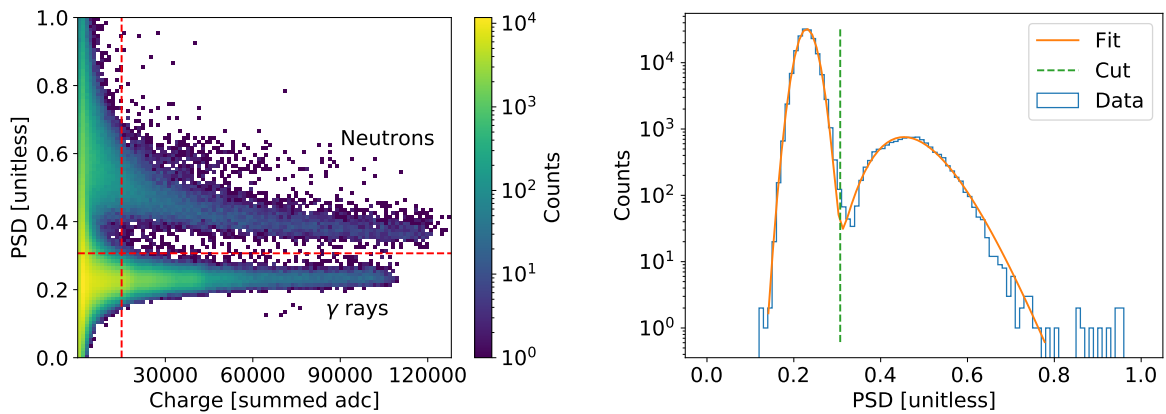

    \centering
    \img{0.45\tw}{analysis/wbls-channel-2-psd-2d.pdf}
    \img{0.45\tw}{analysis/wbls-channel-2-psd-1d.pdf}
    \caption{(Left) PSD metric vs charge for beam events in an example auxiliary detector, showing separation between neutrons and $\gamma$ rays at high charge. (Right) Projection onto the PSD-axis for events above 15000\;ADC units of charge, along with a normal-lognormal fit and subsequently optimized discrimination threshold.}
    \label{fig:psd-example}
\end{figure*}

\subsection{Energy reconstruction}
\label{sec:reconstruction}

    The neutrons produced by 33\;\MeV{} deuterons incident on Be have a broad energy distribution, ranging continuously from the slow spectral region up to 37.4\;\MeV{} (i.e., the incident deuteron energy plus the reaction $Q$-value). While advantageous in allowing simultaneous measurement over a broad energy range, this necessitates event-wise energy reconstruction, which is achieved via two time-of-flight measurements that translate to the neutron energy both before and after interacting with the target scintillator volume. The detection of the scattered neutron in an auxiliary detector establishes a  scattering angle which, for single elastic scatters, kinematically overconstrains the system. For single scatters, the proton energy, $E_p$, is reconstructed in this work using the incident neutron energy, $E_n$, and scattering angle, $\theta$:
\begin{equation}
    \label{eqn:protonE}
    E_p = E_n \sin^2{\theta}.
\end{equation}
    
\begin{figure}
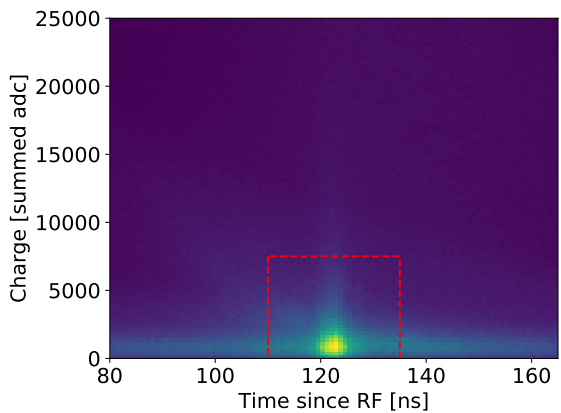

    \centering
    \img{0.45\tw}{analysis/labppo-tof-vs-charge-selection-window.pdf}
    \caption{Charge collected in target PMT vs uncalibrated time since beam extraction during LABPPO data collection. Low energy beam-correlated $\gamma$ rays appear as an isochronic population at low charge. The selection window is illustrated using the red dashed lines.}
    \label{fig:itof-selection}
\end{figure}

    To perform energy reconstruction, the time-of-flight (TOF) measurements are calibrated to correct for cable and system delays. A calibration is performed to determine time differences between interactions in the breakup target and the measurement cell (the ``incoming TOF''), and from the measurement cell to each of the 11 auxiliary detectors (the ``outgoing TOF''). In all cases, the calibration is achieved by selecting on beam-correlated $\gamma$ rays and comparing the measured clock differences to the true TOF given the known speed of light and measured detector positions. Selection of $\gamma$ rays for the outgoing TOF is achieved by exploiting the PSD capabilities of EJ-309, as exemplified in \reffig{psd-example}; $\gamma$-ray selection for the incoming TOF is achieved by selecting low-charge events in the target cell in a given time window, as exemplified in \reffig{itof-selection}. Efforts to apply PSD-based neutron/$\gamma$-ray discrimination using the target scintillators were not fruitful, likely attributable to the dissolved oxygen content. 
    
\begin{figure*}
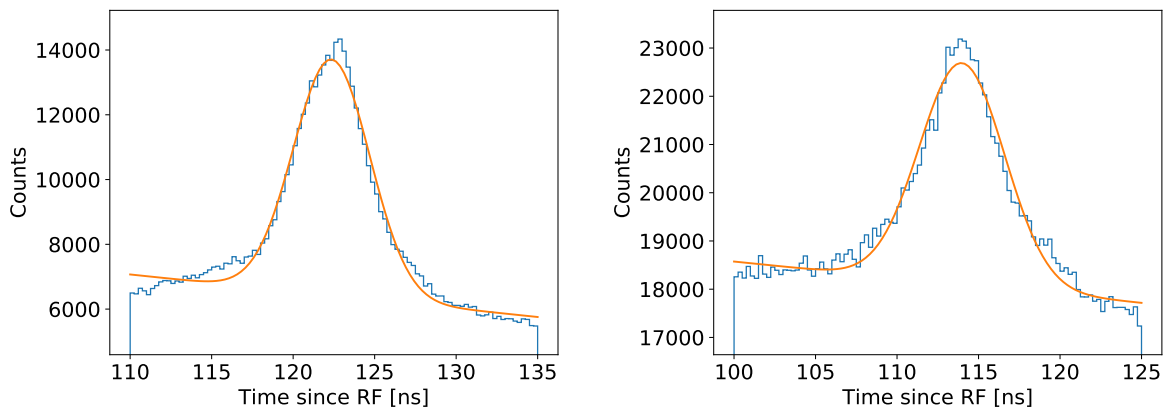

    \centering
     \img{0.45\tw}{analysis/labppo-itof-fit.pdf}
     \img{0.45\tw}{analysis/wbls-itof-fit.pdf}
    \caption{Distribution of measured time differences between the cyclotron RF signal and $\gamma$-ray events in the measurement sample, with empirical fit overlaid, during LABPPO (left) and WbLS (right) data collection.}
    \label{fig:itof-calibration}
\end{figure*}

\begin{figure*}
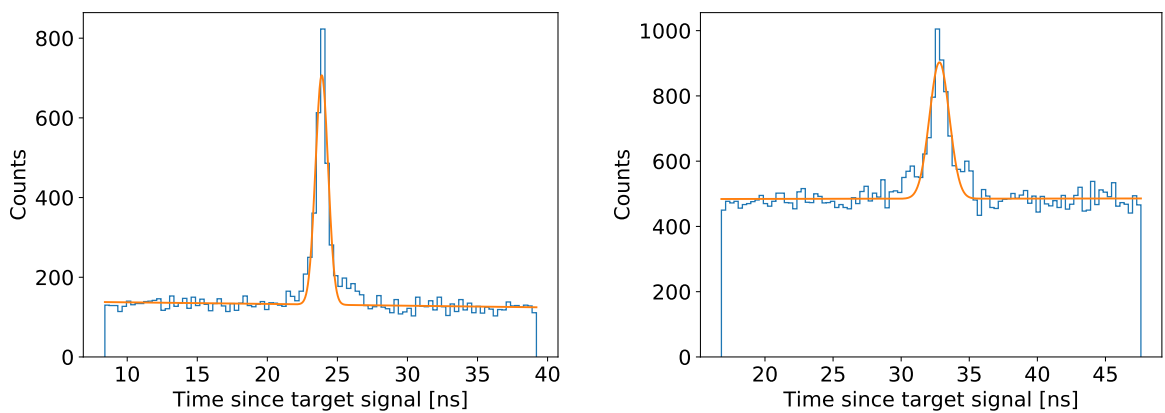

    \centering
     \img{0.45\tw}{analysis/labppo-channel-4-otof-fit.pdf}
     \img{0.45\tw}{analysis/wbls-channel-4-otof-fit.pdf}
    \caption{Distribution of measured time differences between $\gamma$-ray events in the measurement sample and a representative auxiliary detector, with empirical fit overlaid, during LABPPO (left) and WbLS (right) data collection. The data for each material is shown for channel 2, which is located at nominal scattering angles of $80\degree$ and $78\degree$, respectively.}
    \label{fig:otof-calibration}
\end{figure*}

    The resultant distributions of measured $\gamma$-ray time differences are shown in \reffig{itof-calibration} and \reffig{otof-calibration} for the incoming and outgoing TOF, respectively. Each distribution is fit with an empirical function comprised of a Gaussian signal term and a polynomial background term. For the outgoing TOF, the background is modeled using a linear term and is dominated by uncorrelated $\gamma$ rays uniformly distributed in time; a nonzero slope is allowed to account for a potential asymmetry around the $\gamma$-ray population introduced by beam-correlated contamination. For the incoming TOF, there is an additional background of beam-correlated neutrons from previous beam extractions, which have a nontrivial timing structure associated with their energy spectra, and thus a quadratic background term is allowed. The uncertainty in any measured neutron TOF, which propagates to uncertainty in proton recoil energy, is determined both by the uncertainty on the mean of the Gaussian and its width. The width of the incoming TOF is dominated by the temporal profile of the beam pulse. All calibration uncertainties are significantly below 1\%, and the best-fit standard deviations are provided in \reftbl{tof-calibrations}. The relatively poor quality of the fit to the incoming TOF data may be due to the relatively high background rate and shortcomings of the single-Gaussian signal model which in reality is modified by a number of effects, notably the perturbations to the beam due to multiple extraction from the main cyclotron ring. As neutron energy reconstruction is performed under the single beam extraction hypothesis, the relevant quantity for the incoming TOF calibration is the centroid of the $\gamma$-ray population, which is adequately described using the empirical model.
    
\begin{table}
    \begin{tabular}{c c  c  c}
        \- & &
            \multicolumn{2}{c}{
                Standard deviation [ps]
            } \\
        \hline
        \- &
            Channel &
            LABPPO &
            WbLS \\
        \hline
        Incoming &
            - &
            2348.2 &
            2608.8 \\
        Outgoing &
            2 &
            406.6 &
            811.1 \\
        \- &
            3 &
            431.7 &
            914.1 \\
        \- &
            4 &
            448.5 &
            727.9 \\
        \- &
            5 &
            430.3 &
            943.4 \\
        \- &
            6 &
            399.9 &
            896.3 \\
        \- &
            7 &
            420.7 &
            899.5 \\
        \- &
            9 &
            502.8 &
            1019.9 \\
        \- &
            12 &
            403.0 &
            828.8 \\
        \- &
            13 &
            379.9 &
            914.0 \\
        \- &
            14 &
            464.1 &
            730.8 \\
        \- &
            15 &
            423.4 &
            799.9 \\
    \hline
    \end{tabular}
    \caption{Standard deviations of best-fit Gaussian models for TOF distributions of all neutron trajectories, in both the LABPPO and WbLS datasets. Uncertainties on all Gaussian parameters are significantly below 1\%.}
    \label{tbl:tof-calibrations}
\end{table}

    There is ambiguity as to which beam extraction a given neutron detected in the target cell was produced from, associated with the cyclotron operating frequency. The period between beam extractions during data collection was approximately 111\;\nano\second{}. For comparison, the time for a 10\;\MeV{} neutron to travel from the production Be target to the target scintillator cell is approximately 165\;\nano\second{}. A measured incoming TOF can thus be interpreted only as measured modulo the cyclotron period. This ambiguity is resolved by kinematically reconstructing an expected incoming TOF using the outgoing TOF and the known scattering angle. If there is a multiple of the cyclotron period by which the measured and reconstructed incoming TOFs agree to within less than 10\;\nano\second{}, the event is considered kinematically consistent and the ambiguity resolved.

\subsection{Proton light yield extraction}
\label{sec:extraction}

\begin{figure*}
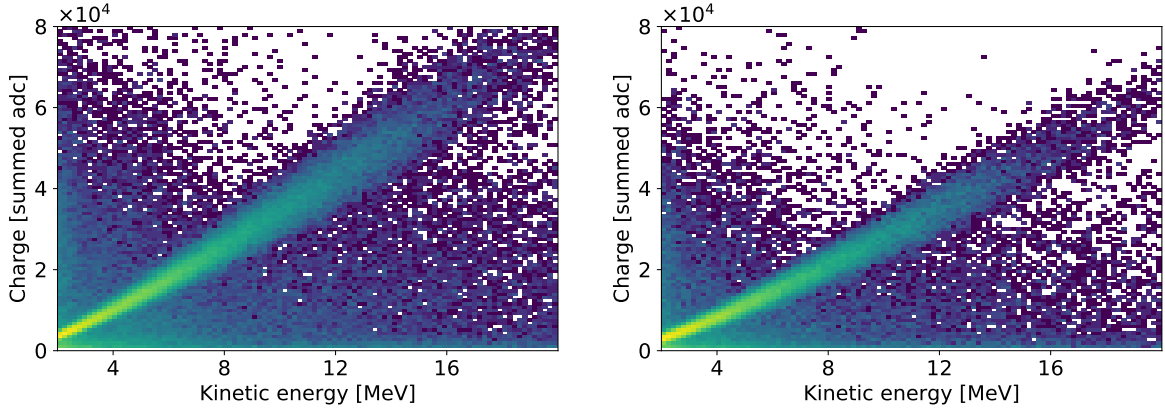

    \centering
    \img{0.45\tw}{analysis/labppo-proton-charge-2d.pdf}
    \img{0.45\tw}{analysis/wbls-proton-charge-2d.pdf}
    \caption{Charge collected in measurement PMT vs energy deposited in the scintillator for kinematically-consistent events in LABPPO (left) and WbLS (right) data.}
    \label{fig:proton-charge-2d}
\end{figure*}

\begin{figure*}
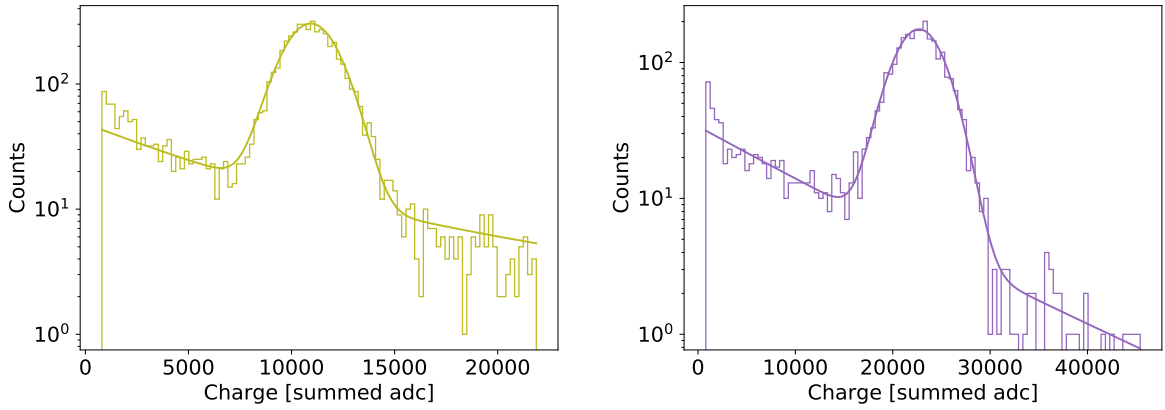

    \centering
    \img{0.45\tw}{analysis/labppo-proton-charge-fit-4-MeV.pdf}
    \img{0.45\tw}{analysis/wbls-proton-charge-fit-8-MeV.pdf}
    \caption{Distributions of charge values for kinematically consistent events, with empirical fits overlaid, of individual proton energy bins: 4.0--4.5\;\MeV{} events in LABPPO (left) and 8.0--9.0\;\MeV{} events in WbLS (right).}
    \label{fig:ply-fits}
\end{figure*}

    Signal events are selected by applying the kinematic consistency criteria described in \refsec{reconstruction} and by selecting neutron events via PID in each auxiliary detector. Two-dimensional distributions of charge and deposited energy for the selected events are shown in \reffig{proton-charge-2d}. To extract the PLY relation, events are partitioned into energy bins, the widths of which are guided by the resolution of single-scatter energy reconstruction, calculated using the TOF calibrations of \refsec{reconstruction} and geometry given in \reftbl{geometry}. A representative charge is assigned to each bin by fitting its population of charge values with an empirical distribution comprised of a Gaussian signal term and two exponential background terms. The centroid of each Gaussian is the representative charge for a given energy bin and, relative to the reference charge defined in \refsec{charge-calibration}, establishes the scale of the relative proton light yield observed. Examples of such fits are shown in \reffig{ply-fits}.

    The fit is formulated using an unbinned maximum likelihood method. Uncertainties are computed using a resampling technique: the statistical uncertainty is computed via bootstrapping \cite{rhinehart2021applied}, wherein the dataset is repeatedly refit under resampling with replacement, and the total uncertainty, which includes systematic effects, is computed similarly, but with analysis parameters which act as sources of uncertainty simultaneously resampled at each iteration. The sources of systematic uncertainty considered include the experimental geometry and timing calibrations: the coordinates of the breakup target, measurement cell, and each auxiliary detector, and the calibration value for each time-of-flight measurement. Each are sampled from normal distributions centered on their nominal values, with standard deviations equal to the associated uncertainties. For each trial, energy reconstruction is performed and each energy bin is refit to extract a representative charge. This procedure generates a non-diagonal covariance matrix due to correlations between energy bins, which stem from the different energy spectra associated with different auxiliary detectors.

\subsection{Modeling ionization quenching}
\label{sec:birks}

    The first model of ionization quenching in organic scintillators was proposed by Birks in 1951 \cite{Birks_1951} and remains widely used in the literature today. For an ion slowing down along a distance $x$ in the scintillating material, the amount of scintillation light produced, $L$, is given by:
\begin{equation}
    \label{eqn:birks}
    \derivative{L}{x} = \frac{S\,\dEdx}{1 + \kB\,\dEdx},
\end{equation}
where $\differential{E}/\differential{x}$ is the stopping power of the ion in the scintillating medium, $S$ establishes the conversion between light produced and energy deposited in the limit of an unquenched system, and $\kB$, termed the Birks constant, introduces nonlinearity characteristic of ionization quenching. Discrepancies have been observed between the Birks model and measured PLY data, particularly at low energy, for a variety of organic scintillators \cite{Yoshida2010,Laplace2022,hong2002scintillation,Williamson1999}. Chou extended the model by introducing a bimolecular quenching term \cite{chou1952nature} which contributes quadratically with the stopping power:
\begin{equation}
    \label{eqn:chou}
    \derivative{L}{x} = \frac{S\,\dEdx}{1 + \kB\,\dEdx + C\pp{\dEdx}^{2}}.
\end{equation}
Using either model, the total photon yield for a fully stopped ion can be found by numerically integrating the quenching relation using a table of stopping powers.

    Quenching parameters are extracted by fitting each model to the measured PLY data via $\chi^{2}$ minimization, with $\chi^{2}$ defined as:
\begin{equation}
    \chi^{2} = \sumon{i,j} \Delta_{i} H_{ij} \Delta_{j},
\end{equation}
where $\Delta_{i}$ is given by:
\begin{equation}
    \Delta_{i} = \pp{Y_{i} - f\pp{E_{i};S,\kB,C}}.
\end{equation}
Here, $E_{i}$ and $Y_{i}$ are the centroid and relative PLY value of the $i^{\text{th}}$ proton energy bin, respectively; $f\pp{E;S,\kB,C}$ denotes the integration of the model up to energy $E$; and $H$ is the inverse of the covariance matrix of the dataset under consideration. Stopping power tables were generated using SRIM \cite{srim}. For table-defined energies $E$, the integral is performed using the trapezoidal rule. For non-table-defined energies, the yield is computed by linearly interpolating between adjacent table-defined yields. Parameter uncertainties and correlations are computed from the covariance matrix.

%% file: inc/results.tex

\section{Results}
\label{sec:results}

\subsection{Proton light yield}

\begin{figure}
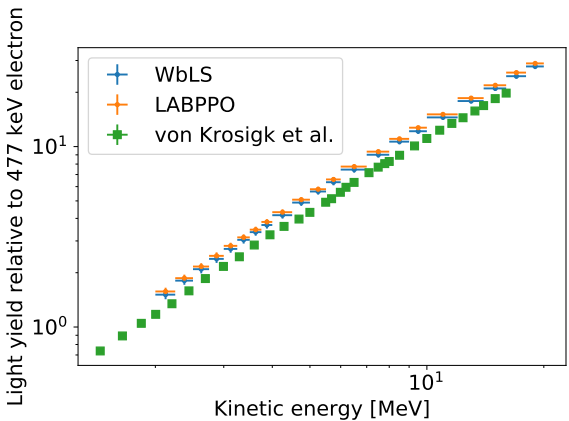

    \centering
    \img{0.45\tw}{results/labppo-wbls-relative-light-yield_loglog.pdf}
    \caption{Proton light yield of LAB + 2\;\gram\per\liter{} PPO and 5\% WbLS, relative to that of a 477\;\keV{} electron. A previous measurement of deoxygenated 2 g/L LABPPO by von Krosigk et al. \cite{vonKrosigk2013} is overlaid.}
    \label{fig:ply-overlay}
\end{figure}

	The light yields of LABPPO and WbLS as a function of proton recoil energy are shown in \reffig{ply-overlay} and listed in \reftbl{labppo-ply} and \reftbl{wbls-ply}, respectively. The horizontal error bars denote the energy bin widths and do not represent uncertainty. The relative PLY of WbLS is consistently lower than that of LABPPO by 3.8\%, although some energy bins below 9.5\;\MeV{} are consistent to within $1\,\sigma$.
	
\begin{table*}
    \centering
    \small
    \begin{tabular}{c | c | c | c | c| c}
        Energy range [\MeV{}] &
                Relative LY &
                Stat. uncertainty [\%] &
                Ref. uncertainty [\%] &
                Total uncertainty [\%] \\
        \hline
        2.00 -- 2.25 &
            1.57 &
            $\pm 0.27$ &
            $\pm 0.84$ &
            $\pm 4.85$ \\
        2.25 -- 2.50 &
            1.87 &
            $\pm 0.25$ &
            $\pm 0.84$ &
            $\pm 4.51$ \\
        2.50 -- 2.75 &
            2.16 &
            $\pm 0.24$ &
            $\pm 0.84$ &
            $\pm 4.32$ \\
        2.75 -- 3.00 &
            2.48 &
            $\pm 0.25$ &
            $\pm 0.84$ &
            $\pm 4.47$ \\
        3.00 -- 3.25 &
            2.82 &
            $\pm 0.24$ &
            $\pm 0.84$ &
            $\pm 4.00$ \\
        3.25 -- 3.50 &
            3.14 &
            $\pm 0.25$ &
            $\pm 0.84$ &
            $\pm 3.89$ \\
        3.50 -- 3.75 &
            3.46 &
            $\pm 0.25$ &
            $\pm 0.84$ &
            $\pm 3.82$ \\
        3.75 -- 4.00 &
            3.82 &
            $\pm 0.26$ &
            $\pm 0.84$ &
            $\pm 3.55$ \\
        4.00 -- 4.50 &
            4.33 &
            $\pm 0.20$ &
            $\pm 0.84$ &
            $\pm 3.60$ \\
        4.50 -- 5.00 &
            5.08 &
            $\pm 0.19$ &
            $\pm 0.84$ &
            $\pm 3.19$ \\
        5.00 -- 5.50 &
            5.80 &
            $\pm 0.19$ &
            $\pm 0.84$ &
            $\pm 3.17$ \\
        5.50 -- 6.00 &
            6.57 &
            $\pm 0.19$ &
            $\pm 0.84$ &
            $\pm 2.96$ \\
        6.00 -- 7.00 &
            7.74 &
            $\pm 0.16$ &
            $\pm 0.84$ &
            $\pm 2.84$ \\
        7.00 -- 8.00 &
            9.38 &
            $\pm 0.17$ &
            $\pm 0.84$ &
            $\pm 2.59$ \\
        8.00 -- 9.00 &
            11.05 &
            $\pm 0.20$ &
            $\pm 0.84$ &
            $\pm 2.68$ \\
        9.00 -- 10.00 &
            12.72 &
            $\pm 0.21$ &
            $\pm 0.84$ &
            $\pm 2.27$ \\
        10.00 -- 12.00 &
            15.09 &
            $\pm 0.19$ &
            $\pm 0.84$ &
            $\pm 1.99$ \\
        12.00 -- 14.00 &
            18.56 &
            $\pm 0.25$ &
            $\pm 0.84$ &
            $\pm 1.91$ \\
        14.00 -- 16.00 &
            21.86 &
            $\pm 0.34$ &
            $\pm 0.84$ &
            $\pm 1.89$ \\
        16.00 -- 18.00 &
            25.71 &
            $\pm 0.49$ &
            $\pm 0.84$ &
            $\pm 1.87$ \\
        18.00 -- 20.00 &
            28.84 &
            $\pm 0.76$ &
            $\pm 0.84$ &
            $\pm 1.75$ \\
    \end{tabular}
    \caption{Light yield of proton recoils, relative to that of a 477\;\keV{} electron, in LAB + 2\;\gram\per\liter{} PPO, and associated uncertainties (from left to right): statistical uncertainty, uncertainty on reference charge, and total uncertainty including systematic effects. A correlation matrix of the per-bin uncertainties is available upon request.}
    \label{tbl:labppo-ply}
\end{table*}

\begin{table*}
    \centering
    \small
    \begin{tabular}{c | c | c | c | c| c}
        Energy range [\MeV{}] &
                Relative LY &
                Stat. uncertainty [\%] &
                Ref. uncertainty [\%] &
                Total uncertainty [\%] \\
        \hline
        2.00 -- 2.25 &
            1.51 &
            $\pm 0.58$ &
            $\pm 2.54$ &
            $\pm 5.68$ \\
        2.25 -- 2.50 &
            1.81 &
            $\pm 0.45$ &
            $\pm 2.54$ &
            $\pm 5.15$ \\
        2.50 -- 2.75 &
            2.10 &
            $\pm 0.43$ &
            $\pm 2.54$ &
            $\pm 5.23$ \\
        2.75 -- 3.00 &
            2.39 &
            $\pm 0.40$ &
            $\pm 2.54$ &
            $\pm 5.05$ \\
        3.00 -- 3.25 &
            2.71 &
            $\pm 0.37$ &
            $\pm 2.54$ &
            $\pm 4.78$ \\
        3.25 -- 3.50 &
            3.04 &
            $\pm 0.37$ &
            $\pm 2.54$ &
            $\pm 4.66$ \\
        3.50 -- 3.75 &
            3.36 &
            $\pm 0.40$ &
            $\pm 2.54$ &
            $\pm 4.49$ \\
        3.75 -- 4.00 &
            3.68 &
            $\pm 0.38$ &
            $\pm 2.54$ &
            $\pm 4.45$ \\
        4.00 -- 4.50 &
            4.17 &
            $\pm 0.28$ &
            $\pm 2.54$ &
            $\pm 4.25$ \\
        4.50 -- 5.00 &
            4.90 &
            $\pm 0.29$ &
            $\pm 2.54$ &
            $\pm 4.18$ \\
        5.00 -- 5.50 &
            5.64 &
            $\pm 0.28$ &
            $\pm 2.54$ &
            $\pm 3.92$ \\
        5.50 -- 6.00 &
            6.34 &
            $\pm 0.28$ &
            $\pm 2.54$ &
            $\pm 3.90$ \\
        6.00 -- 7.00 &
            7.45 &
            $\pm 0.24$ &
            $\pm 2.54$ &
            $\pm 3.88$ \\
        7.00 -- 8.00 &
            9.01 &
            $\pm 0.24$ &
            $\pm 2.54$ &
            $\pm 3.60$ \\
        8.00 -- 9.00 &
            10.65 &
            $\pm 0.25$ &
            $\pm 2.54$ &
            $\pm 3.35$ \\
        9.00 -- 10.00 &
            12.17 &
            $\pm 0.27$ &
            $\pm 2.54$ &
            $\pm 3.39$ \\
        10.00 -- 12.00 &
            14.51 &
            $\pm 0.26$ &
            $\pm 2.54$ &
            $\pm 3.10$ \\
        12.00 -- 14.00 &
            17.87 &
            $\pm 0.32$ &
            $\pm 2.54$ &
            $\pm 3.01$ \\
        14.00 -- 16.00 &
            21.00 &
            $\pm 0.39$ &
            $\pm 2.54$ &
            $\pm 3.14$ \\
        16.00 -- 18.00 &
            24.54 &
            $\pm 0.64$ &
            $\pm 2.54$ &
            $\pm 3.05$ \\
        18.00 -- 20.00 &
            27.89 &
            $\pm 0.90$ &
            $\pm 2.54$ &
            $\pm 3.14$ \\
    \end{tabular}
    \caption{Light yield of proton recoils, relative to a 477\;\keV{} electron, in 5\% WbLS, and associated uncertainties (from left to right): statistical uncertainty, uncertainty on reference charge, and total uncertainty including systematic effects. A correlation matrix of the per-bin uncertainties is available upon request.}
    \label{tbl:wbls-ply}
\end{table*}

    Previous PLY measurements of several LABPPO formulations were performed by von Krosigk et al.\ using a neutron beam at the Physikalisch-Technische Bundesanstalt (PTB)~\cite{vonKrosigk2013}. The PTB measurement for a deoxygenated 2~g/L LABPPO scintillator is also shown in \reffig{ply-overlay}. The relative PLY data are systematically lower than the LABPPO PLY obtained in this work by $15-20$\%. A discrepancy between the PLY of the two samples is not unexpected as the LABPPO measured at PTB was deoxygenated via bubbling with gaseous argon, which removes molecular oxygen, whereas the sample measured in this work was not. Such deoxygenation has been shown to impact ionization quenching~\cite{Berlman1961,Birks}, though the relative proton light yield would be expected to decrease in aerated samples, not increase as is observed here, due to the differential impact of oxygen quenching of triplet states given the higher fraction of delayed light for proton recoils relative to electrons. 

    There are a number of factors that can potentially explain this discrepancy. Different integration lengths used in waveform processing can lead to significant discrepancies in relative proton light yields~\cite{Brown2018,Laplace2020-Bias}. This is due to differences in the scintillation temporal profiles of electrons and protons, as well as potential variation in the proton pulse shape with recoil energy: use of an integration length that is too short results in a pulse integral that is not proportional to the total number of scintillation photons. The integration length used in this work is 140\;\nano\second{}, which was chosen to ensure that $>95$\% of the light was collected. The integration length used in the PTB measurement is not reported in \cite{vonKrosigk2013}. 
    
    The reference charge calibration also represents a potential source of bias. The electron light yield of LABPPO has been shown to deviate from linearity below $\sim 400$~keV \cite{WanChanTseung2011}. The PTB group used multiple $\gamma$-ray sources but assumed electron light linearity, equivalent to fixing the offset parameter $b=0$. For LABPPO, the multi-source calibration performed in this work leads to an offset parameter, $b=34.7\pm1.2$~keV, indicative of electron light nonlinearity. The average charge per unit energy can be calculated for the single Compton edge fits described in \refsec{charge-calibration}. This charge per unit energy is 5.2\% greater when using the 1547\;\keV{} Compton edge from \isotope{Bi}{207} compared to the 477\;\keV{} Compton edge from \isotope{Cs}{137}; this value is 6.6\% greater if the 2520\;\keV{} Compton edge from \isotope{Na}{24} is used.

    Finally, the edge characterization method employed in \cite{vonKrosigk2013} to extract the PLY is known to be subject to bias \cite{Brown2018,Weldon2020}. In particular, the importance of neutron response modeling to the PTB measurement necessitates the need to extrapolate the light yield curve to lower energies in order to properly account for multiple neutron scatters, whereas the kinematic consistency and signal extraction methods employed in this work are model independent.

\subsection{Model compatibility}

\begin{figure*}
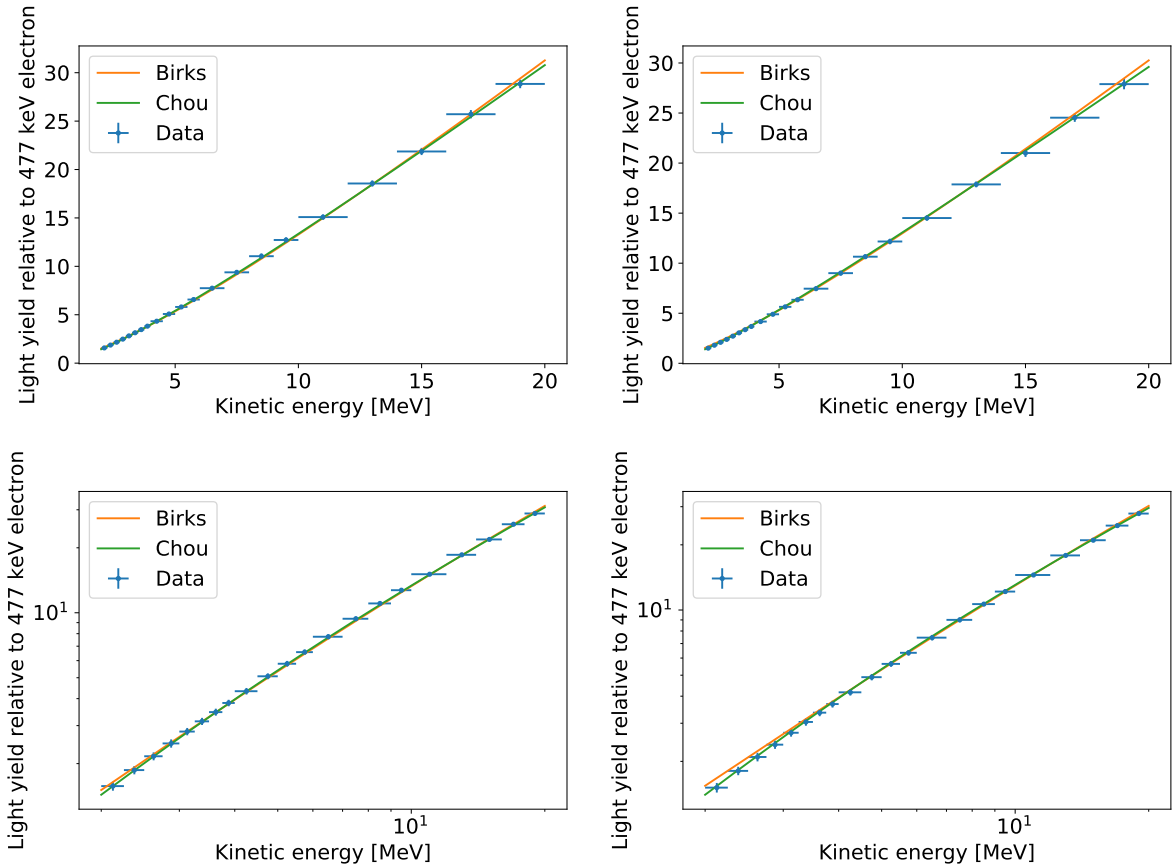

    \centering
     \img{0.45\tw}{results/labppo-model-fits.pdf}
     \img{0.45\tw}{results/wbls-model-fits.pdf}
     \img{0.45\tw}{results/labppo-model-fits_loglog.pdf}
     \img{0.45\tw}{results/wbls-model-fits_loglog.pdf}
    \caption{Best-fit quenching models compared to the measured PLY for LABPPO (left) and WbLS (right), shown with both linear (top) and logarithmic (bottom) axes. The Birks model fails to reproduce the low-energy behavior of both materials, which is better modeled with the inclusion of the Chou bimolecular quenching term.}
    \label{fig:ply-model-fits}
\end{figure*}

    Figure~\ref{fig:ply-model-fits} shows the best-fit quenching models for the LABPPO and WbLS relative proton light yield data obtained using the Birks and Chou parameterizations (see Eqs.~\ref{eqn:birks} and~\ref{eqn:chou}, respectively). The best-fit model parameters are listed in \reftbl{birks_float}. The Chou model provides a better fit for each material and significant deviations are observed for the Birks fit of the WbLS data below 3\;\MeV{} proton recoil energy. The parameter correlation between $S$ and $kB$ in the Birks model is 87.2\% and 87.4\% for the LABPPO and WbLS datasets, respectively. Correlation matrices associated with the Chou model are provided in \reftbl{covariance}.
    
\begin{table*}
    \centering
    \begin{tabular}{c  c  c  c  c}
    \hline
        \- &
            \multicolumn{2}{c}{
                Birks
            } &
            \multicolumn{2}{c}{
                Chou
            } \\
        \- &
            LABPPO &
            WbLS &
            LABPPO &
            WbLS \\
        \hline
        $S\;[\MeV^{-1}]$ &
            $2.193\pm{0.053}$ &
            $2.082\pm{0.071}$ &
            $1.963\pm{0.074}$ &
            $1.776\pm{0.079}$ \\
        $\kB\;[\centi\meter\per\GeV]$ &
            $7.08\pm{0.45}$ &
            $5.95\pm{0.43}$ &
            $3.76\pm{0.91}$ &
            $1.65\pm{0.81}$ \\
        $C\;[\centi\meter^{2}\per\GeV^{2}]$ &
            - &
            - &
            $9.88\pm{2.74}$ &
            $13.30\pm{2.70}$ \\
        $\chi^{2}/\ndf$ &
            $36.6/19$ &
            $44.7/19$ &
            $22.8/18$ &
            $17.3/18$ \\
            \hline
    \end{tabular}
    \caption{Best-fit model parameters for the LABPPO and WbLS proton light yields, relative to that of a $477\;\keV{}$ electron. Neither material is well modeled using the Birks formalism, but both are adequately described using the Chou model.}
    \label{tbl:birks_float}
\end{table*}

\begin{table}
    \centering
    \small
    \begin{tabular}{c | c c c}
        LABPPO &
            $S$ &
            $\kB$ &
            $C$ \\
        \hline
        $S$ &
            100.0 &
            93.9 &
            -73.5 \\
        $\kB$ &
            \- &
            100.0 &
            -87.0 \\
        $C$ &
            \- &
            \- &
            100.0 \\
        \-
            \- &
            \- &
            \- \\
        WbLS &
            $S$ &
            $\kB$ &
            $C$ \\
        \hline
        $S$ &
            100.0 &
            93.7 &
            -70.2 \\
        $\kB$ &
            \- &
            100.0 &
            -84.9 \\
        $C$ &
            \- &
            \- &
            100.0 \\
    \end{tabular}
    \caption{Correlation matrices of parameter uncertainties for the Chou quenching model, in units of percent.}
    \label{tbl:covariance}
\end{table}

    The PLY of LAB + 2\;\gram\per\liter{} PPO (+ 15\;\milli\gram{}\per\liter{} bis-MSB, a secondary fluor) were measured using a proton beam at the NASA Space Radiation Laboratory at Brookhaven National Laboratory, and fit with Birks' law in \cite{bignell2015jinst}. The reported best-fit Birks' constant of $\kB = \pp{7.0 \pm 0.1}\;\centi\meter\per\GeV$) is consistent with the present result, although it should be noted that the Brookhaven measurement was performed at energies above 20\;\MeV{}.
    The PTB study investigated ionization quenching in LABPPO using the Chou model \cite{vonKrosigk2013}. In that work, scintillation light was quantified using an electron-equivalent energy in units of $\MeVee\per\MeV{}$, and $S$ was fixed to a value of 1 $\MeVee\per\MeV{}$. In this work, scintillation light was determined relative to that produced by a 477\;\keV{} electron, which gives a value of $S = \pp{477\;\keV}^{-1} = 2.095\;\MeV^{-1}$ in the absence of electron light quenching. The best-fit model in the PTB study was consistent with a quadratic coefficient, $C$, of zero, i.e., equivalent to the model provided in \refeqn{birks}, although a metric directly quantifying the goodness-of-fit was not reported. In contrast, in this work, a nonzero quadratic coefficient is preferred. The Birks constant extracted in the PTB study, $\kB = 9.8\;\centi\meter\per\GeV$, is larger than that found in this work, while the fixed value of $S$ is smaller (though consistent to within $1\,\sigma$). As the $S$ and $kB$ parameter errors are positively correlated, a decrease in the estimate of $S$ would result in a decreased estimate of $kB$ for the same predicted light yield. Hence, fixing $S = 2.095\;\MeV^{-1}$ in for this work would result in a smaller value of $kB$ for the Birks fit, representing an even larger discrepancy with the PTB quenching parameter.
    
    The PTB measurement extended a few hundred \keV{} below the 2\;\MeV{} floor used in this work, but the best-fit model failed in the high energy region, systematically predicting an excess light yield above 12\;\MeV{}. Additional PLY measurements, particularly at lower energy and with deoxygenated samples, would help resolve tension with the PTB study.

\subsection{Discussion}

    In \cite{vonKrosigk2013}, the effect of different levels of proton quenching on the detection of supernova neutrinos in large LAB-based detectors is discussed. An effective detection threshold corresponding to a 200\;\keV{} kinetic energy electron is assumed, driven by the 156~keV endpoint energy of \isotope{C}{14} $\beta$-decay, a prominent background internal to organic liquid scintillators. The relatively high degree of ionization quenching determined in that work (quantified by $\kB$) was found to reduce the event rate in a SNO+-like detector \cite{snoplus_detector} by 16\% when compared to a reference quenching parameter of $7.3\;\centi\meter\per\GeV$ taken from \cite{Beacom2011}. The Birks' constant extracted in this work, $\kB = \pp{7.08 \pm{0.45}}\;\centi\meter\per\GeV$, is consistent with the reference value, which restores a relatively optimistic outlook for supernova neutrino detection. Furthermore, the uniform excess in relative PLY shown in \reffig{ply-overlay} translates to an increase in proton energy resolution, which would allow for a lower detection threshold and higher detection rate.

%% file: inc/conclusion.tex

\section{Conclusion}
\label{sec:conclusion}

    The PLY of LAB with 2\;\gram\per\liter{} PPO and 5\% WbLS were measured using a double time-of-flight technique at the 88-Inch Cyclotron at Lawrence Berkeley National Laboratory. The results obtained in this work for LABPPO exposed to the atmosphere disagree with a previous measurement of deoxygenated LABPPO performed at PTB \cite{vonKrosigk2013}. Additional measurements using both oxygenated and deoxygenated samples, ideally extending to lower proton energies, would help to resolve this discrepancy. Application of ionization quenching models revealed that neither material is adequately modeled using the Birks relation and inclusion of a bimolecular quenching term in the manner of Chou was required. These results are relevant to the design of future WbLS applications involving the detection of neutrons and protons. In the context of neutrino physics, this includes the discrimination of fast neutrons from electron-like coincidence signals, e.g., IBD events, and potential measurements of the flavor-inclusive energy spectra of neutrinos from future supernovae.

%% file: inc/acknowledgements.tex

\begin{acknowledgements}

    The authors thank the 88-Inch Cyclotron operations and facilities staff for their help in performing these experiments. This work was performed under the auspices of the U.S. Department of Energy by Lawrence Berkeley National Laboratory under Contract DE-AC02-05CH11231. The project was funded by the U.S. Department of Energy, National Nuclear Security Administration, Office of Defense Nuclear Nonproliferation Research and Development (DNN R\&D). This material is based upon work supported in part by the U.S.\ Department of Energy National Nuclear Security Administration through the Nuclear Science and Security Consortium under Award DE-NA0003180. EJC was funded by the Consortium for Monitoring, Technology, and Verification under Department of Energy National Nuclear Security Administration award number DE-NA0003920.

\end{acknowledgements}

%% file: paper.bbl
\begin{thebibliography}{10}
\providecommand{\url}[1]{{#1}}
\providecommand{\urlprefix}{URL }
\expandafter\ifx\csname urlstyle\endcsname\relax
  \providecommand{\doi}[1]{DOI \discretionary{}{}{}#1}\else
  \providecommand{\doi}{DOI \discretionary{}{}{}\begingroup
  \urlstyle{rm}\Url}\fi

\bibitem{Reines1956}
F.~Reines, C.L. Cowan, Nature \textbf{178}(4531), 446 (1956).
\newblock \doi{10.1038/178446a0}

\bibitem{An2012}
F.P. An, et~al., Phys. Rev. Lett. \textbf{108}, 171803 (2012).
\newblock \doi{10.1103/PhysRevLett.108.171803}.
\newblock
  \urlprefix\url{https://link.aps.org/doi/10.1103/PhysRevLett.108.171803}

\bibitem{Andringa2016}
S.~Andringa, et~al., Adv. High Energy Phys. \textbf{2016}, 1 (2016).
\newblock \doi{10.1155/2016/6194250}.
\newblock \urlprefix\url{http://dx.doi.org/10.1155/2016/6194250}

\bibitem{Yeh2011}
M.~Yeh, S.~Hans, W.~Beriguete, R.~Rosero, L.~Hu, R.L. Hahn, M.V. Diwan, D.E.
  Jaffe, S.H. Kettell, L.~Littenberg, Nuclear Instruments and Methods in
  Physics Research Section A: Accelerators, Spectrometers, Detectors and
  Associated Equipment \textbf{660}(1), 51 (2011).
\newblock \doi{https://doi.org/10.1016/j.nima.2011.08.040}.
\newblock
  \urlprefix\url{https://www.sciencedirect.com/science/article/pii/S0168900211016615}

\bibitem{ANNIE}
A.R. Back, et~al., arXiv:1707.08222 [physics.ins-det]  (2017)

\bibitem{AITNEO}
M.~Askins, et~al.,   (2015)

\bibitem{Askins2020}
M.~Askins, Z.~Bagdasarian, N.~Barros, E.W. Beier, E.~Blucher, R.~Bonventre,
  E.~Bourret, E.J. Callaghan, J.~Caravaca, M.~Diwan, et~al., The European
  Physical Journal C \textbf{80}(5), 416 (2020).
\newblock \doi{10.1140/epjc/s10052-020-7977-8}

\bibitem{Bignell2015}
L.J. Bignell, D.~Beznosko, M.V. Diwan, S.~Hans, D.E. Jaffe, S.~Kettell,
  R.~Rosero, H.W. Themann, B.~Viren, E.~Worcester, M.~Yeh, C.~Zhang, Journal of
  Instrumentation \textbf{10}(12), P12009 (2015).
\newblock \doi{10.1088/1748-0221/10/12/p12009}.
\newblock \urlprefix\url{https://doi.org/10.1088/1748-0221/10/12/p12009}

\bibitem{Onken2020}
D.R. Onken, F.~Moretti, J.~Caravaca, M.~Yeh, G.D. Orebi~Gann, E.D. Bourret,
  Materials Advances \textbf{1}, 71 (2020).
\newblock \doi{10.1039/D0MA00055H}.
\newblock \urlprefix\url{http://dx.doi.org/10.1039/D0MA00055H}

\bibitem{Caravaca2020}
J.~Caravaca, B.J. Land, M.~Yeh, G.D. Orebi~Gann, The European Physical Journal
  C \textbf{80}(9), 867 (2020).
\newblock \doi{10.1140/epjc/s10052-020-8418-4}

\bibitem{Land2021}
B.J. Land, Z.~Bagdasarian, J.~Caravaca, M.~Smiley, M.~Yeh, G.D. Orebi~Gann,
  Phys. Rev. D \textbf{103}(5), 052004 (2021).
\newblock \doi{10.1103/PhysRevD.103.052004}

\bibitem{Bernstein2020}
A.~Bernstein, N.~Bowden, B.L. Goldblum, P.~Huber, I.~Jovanovic, J.~Mattingly,
  Rev. Mod. Phys. \textbf{92}, 011003 (2020).
\newblock \doi{10.1103/RevModPhys.92.011003}.
\newblock \urlprefix\url{https://link.aps.org/doi/10.1103/RevModPhys.92.011003}

\bibitem{Yoshida2010}
S.~Yoshida, T.~Ebihara, T.~Yano, A.~Kozlov, T.~Kishimoto, I.~Ogawa, R.~Hazama,
  S.~Umehara, K.~Mukaida, K.~Ichihara, Y.~Hirano, I.~Murata, J.~Datemichi,
  H.~Sugimoto, Nuclear Instruments and Methods in Physics Research Section A:
  Accelerators, Spectrometers, Detectors and Associated Equipment
  \textbf{622}(3), 574 (2010).
\newblock \doi{https://doi.org/10.1016/j.nima.2010.07.087}.
\newblock
  \urlprefix\url{https://www.sciencedirect.com/science/article/pii/S0168900210017018}

\bibitem{Harrig2018}
K.P. Harrig, B.L. Goldblum, J.A. Brown, D.L. Bleuel, L.A. Bernstein, J.~Bevins,
  M.~Harasty, T.A. Laplace, E.F. Matthews, Nuclear Instruments and Methods in
  Physics Research Section A: Accelerators, Spectrometers, Detectors and
  Associated Equipment \textbf{877}, 359  (2018).
\newblock \doi{https://doi.org/10.1016/j.nima.2017.09.051}.
\newblock
  \urlprefix\url{http://www.sciencedirect.com/science/article/pii/S0168900217310215}

\bibitem{EJ309}
Eljen Technology, \emph{{Neutron/Gamma PSD Liquid Scintillator EJ-301, EJ-309}}
  (2021).
\newblock
  \urlprefix\url{https://eljentechnology.com/images/products/data_sheets/EJ-301_EJ-309.pdf}

\bibitem{Brown2018}
J.A. Brown, B.L. Goldblum, T.A. Laplace, K.P. Harrig, L.A. Bernstein, D.L.
  Bleuel, W.~Younes, D.~Reyna, E.~Brubaker, P.~Marleau, Journal of Applied
  Physics \textbf{124}(4), 045101 (2018).
\newblock \doi{10.1063/1.5039632}

\bibitem{BrownThesis}
J.A. Brown, A double time of flight method for measuring proton light yield.
\newblock Ph.D. thesis, University of California, Berkeley (2017)

\bibitem{Friend2011}
M.~Friend, G.~Franklin, B.~Quinn, Nucl. Instrum. Methods Phys. Res. A
  \textbf{676} (2011)

\bibitem{LEDspec}
{OSRAM LED} Engin, \emph{High Efficiency {VIOLET LED} Emitter LZ1-00UB00}
  (2018).
\newblock
  \urlprefix\url{https://dammedia.osram.info/im/bin/osram-dam-5412887/LED\%20Engin_Datasheet_LuxiGen_LZ1-00UB00_rev2.1_20181120.pdf}

\bibitem{GEANT4}
S.~Agostinelli, et~al., Nucl. Instrum. Meth. A \textbf{506}, 250 (2003).
\newblock \doi{10.1016/S0168-9002(03)01368-8}

\bibitem{Dietze1982}
G.~Dietze, H.~Klein, Nuclear Instruments and Methods in Physics Research
  \textbf{193}(3), 549  (1982).
\newblock \doi{https://doi.org/10.1016/0029-554X(82)90249-X}

\bibitem{Laplace2020-EJ20x}
T.A. Laplace, B.L. Goldblum, J.A. Brown, D.L. Bleuel, C.A. Brand, G.~Gabella,
  T.~Jordan, C.~Moore, N.~Munshi, Z.W. Sweger, A.~Sweet, E.~Brubaker, Nucl.
  Instrum. Methods Phys. Res. A \textbf{954}, 161444 (2020).
\newblock \doi{https://doi.org/10.1016/j.nima.2018.10.122}.
\newblock
  \urlprefix\url{http://www.sciencedirect.com/science/article/pii/S0168900218314360}.
\newblock {Symposium on Radiation Measurements and Applications XVII}

\bibitem{ROOT}
R.~Brun, F.~Rademakers, Nucl. Instrum. Methods Phys. Res. A \textbf{389}(1), 81
   (1997).
\newblock \doi{http://dx.doi.org/10.1016/S0168-9002(97)00048-X}.
\newblock
  \urlprefix\url{http://www.sciencedirect.com/science/article/pii/S016890029700048X}

\bibitem{rhinehart2021applied}
R.R. Rhinehart, R.M. Bethea, \emph{Applied Engineering Statistics} (CRC Press,
  2021)

\bibitem{Birks_1951}
J.B. Birks, Proceedings of the Physical Society. Section A \textbf{64}(10), 874
  (1951).
\newblock \doi{10.1088/0370-1298/64/10/303}.
\newblock \urlprefix\url{https://doi.org/10.1088/0370-1298/64/10/303}

\bibitem{Laplace2022}
T.A. Laplace, B.L. Goldblum, J.A. Brown, G.~LeBlanc, T.~Li, J.J. Manfredi,
  E.~Brubaker, Materials Advances \textbf{3}, 5871 (2022).
\newblock \doi{10.1039/D2MA00388K}

\bibitem{hong2002scintillation}
J.~Hong, W.~Craig, P.~Graham, C.~Hailey, N.~Spooner, D.~Tovey, Astroparticle
  Physics \textbf{16}(3), 333 (2002)

\bibitem{Williamson1999}
J.F. Williamson, J.F. Dempsey, A.S. Kirov, J.I. Monroe, W.R. Binns,
  H.~Hedtjärn, Physics in Medicine and Biology \textbf{44}(4), 857 (1999).
\newblock \doi{10.1088/0031-9155/44/4/004}.
\newblock \urlprefix\url{https://doi.org/10.1088/0031-9155/44/4/004}

\bibitem{chou1952nature}
C.~Chou, Physical Review \textbf{87}(5), 904 (1952)

\bibitem{srim}
{J. F. Ziegler}.
\newblock {Computer program, available at \url{http://srim.org}. Accessed
  September 2021.}

\bibitem{vonKrosigk2013}
B.~von Krosigk, L.~Neumann, R.~Nolte, S.~R\"{o}ttger, K.~Zuber, The European
  Physical Journal C \textbf{73}(4) (2013).
\newblock \doi{10.1140/epjc/s10052-013-2390-1}.
\newblock \urlprefix\url{http://dx.doi.org/10.1140/epjc/s10052-013-2390-1}

\bibitem{Berlman1961}
I.B. Berlman, The Journal of Chemical Physics \textbf{34}(2), 598 (1961).
\newblock \doi{10.1063/1.1700992}.
\newblock \urlprefix\url{https://doi.org/10.1063/1.1700992}

\bibitem{Birks}
J.~Birks, \emph{The Theory and Practice of Scintillation Counting}.
\newblock International Series of Monographs in Electronics and Instrumentation
  (Pergamon Press, New York, 1964)

\bibitem{Laplace2020-Bias}
T.A. Laplace, B.L. Goldblum, J.A. Brown, J.J. Manfredi, Nuclear Instruments and
  Methods in Physics Research Section A: Accelerators, Spectrometers, Detectors
  and Associated Equipment \textbf{959}, 163485 (2020).
\newblock \doi{10.1016/j.nima.2020.163485}.
\newblock \urlprefix\url{http://dx.doi.org/10.1016/j.nima.2020.163485}

\bibitem{WanChanTseung2011}
H.~{Wan Chan Tseung}, J.~Kaspar, N.~Tolich, Nuclear Instruments and Methods in
  Physics Research Section A: Accelerators, Spectrometers, Detectors and
  Associated Equipment \textbf{654}(1), 318 (2011).
\newblock \doi{https://doi.org/10.1016/j.nima.2011.06.095}.
\newblock
  \urlprefix\url{https://www.sciencedirect.com/science/article/pii/S0168900211013908}

\bibitem{Weldon2020}
R.~Weldon, J.~Mueller, P.~Barbeau, J.~Mattingly, Nuclear Instruments and
  Methods in Physics Research Section A: Accelerators, Spectrometers, Detectors
  and Associated Equipment \textbf{953}, 163192 (2020).
\newblock \doi{https://doi.org/10.1016/j.nima.2019.163192}.
\newblock
  \urlprefix\url{https://www.sciencedirect.com/science/article/pii/S0168900219314871}

\bibitem{bignell2015jinst}
L.J. Bignell, D.~Beznosko, M.~Diwan, S.~Hans, D.~Jaffe, S.~Kettell, R.~Rosero,
  H.~Themann, B.~Viren, E.~Worcester, M.~Yeh, C.~Zhang, Journal of
  Instrumentation \textbf{10}(12), 12009 (2015)

\bibitem{snoplus_detector}
V.~Albanese, R.~Alves, M.~Anderson, S.~Andringa, L.~Anselmo, E.~Arushanova,
  S.~Asahi, M.~Askins, D.~Auty, A.~Back, et~al., Journal of Instrumentation
  \textbf{16}(08), P08059 (2021).
\newblock \doi{10.1088/1748-0221/16/08/p08059}.
\newblock \urlprefix\url{http://dx.doi.org/10.1088/1748-0221/16/08/P08059}

\bibitem{Beacom2011}
B.~Dasgupta, J.F. Beacom, Phys. Rev. D \textbf{83}, 113006 (2011).
\newblock \doi{10.1103/PhysRevD.83.113006}.
\newblock \urlprefix\url{https://link.aps.org/doi/10.1103/PhysRevD.83.113006}

\end{thebibliography}
